\documentclass[runningheads]{llncs}
\usepackage{color}
\usepackage{graphicx}
\usepackage{multirow}
\usepackage{enumitem}
\usepackage{amsmath}
\usepackage{amsfonts}

\begin{document}
\mainmatter

\title{Adversarial Learning with Multiscale Features and Kernel Factorization for Retinal Blood Vessel Segmentation}

\author{Farhan Akram\inst{1,\thanks{Corresponding Author: farhana@bii.a-star.edu.sg}} \and Vivek Kumar Singh\inst{2} \and Hatem A. Rashwan\inst{2} \and Mohamed Abdel-Nasser\inst{2} \and  Md. Mostafa Kamal Sarker\inst{2} \and Nidhi Pandey\inst{3} \and Domenec Puig\inst{2}}
\titlerunning{Adversarial Learning with Multiscale Features and Kernel Factorization for Retinal Blood Vessel Segmentation}
%
\authorrunning{F. Akram et al.}	
\institute{Imaging Informatics Division, Bioinformatics Institute, Singapore. \and DEIM, Universitat Rovira i Virgili, Spain. \and
Department of Medicine and Health Sciences, Universitat Rovira i Virgili, Spain.\\
}
%
\maketitle              
\vspace{-2mm}
\begin{abstract}

In this paper, we propose an efficient blood vessel segmentation method for the eye fundus images using adversarial learning with multiscale features and kernel factorization. In the generator network of the adversarial framework, spatial pyramid pooling, kernel factorization and squeeze excitation block are employed to enhance the feature representation in spatial domain on different scales with reduced computational complexity. In turn, the discriminator network of the adversarial framework is formulated by combining convolutional layers with an additional squeeze excitation block to differentiate the generated segmentation mask from its respective ground truth. Before feeding the images to the network, we pre-processed them by using edge sharpening and Gaussian regularization to reach an optimized solution for vessel segmentation. The output of the trained model is post-processed using morphological operations to remove the small speckles of noise. The proposed method qualitatively and quantitatively outperforms state-of-the-art vessel segmentation methods using DRIVE and STARE datasets.

\vspace{-2mm}
\keywords{Retinal Vessel Segmentation, Spatial Pyramid Pooling, Kernel Factorization, Squeeze Excitation.}
\end{abstract}
\section{Introduction}
\vspace{-2mm}
The segmentation of retinal blood vessels in the fundus photographs plays a critical role in the medical diagnosis, screening, and treatment of several ophthalmologic diseases \cite{macgillivray2014retinal}. The study of variations in vessel morphology such as shape, tortuosity, branching pattern and width can help in an early diagnosis in which an accurate segmentation is needed. Numerous methods \cite{vostatek2017performance} have been proposed in the literature, which yields a similar segmentation accuracy compared to the trained ophthalmologists. However, there are still some open challenges which are needed to be addressed. That includes vessel segmentation in the presence of intensity inhomogeneity, segmentation of vessels near the bifurcation and crossover regions and lastly the segmentation of thin vessel structures. 

Since deep learning is evolving at a vast speed, many deep learning based methods are recently proposed for the blood vessel segmentation in the fundus photographs \cite{oliveira2018retinal}. A multiscale convolutional neural network (CNN) architecture and fully connected conditional random fields (CRFs) with an improved cross-entropy loss function was recently proposed recently  in~\cite{hu2018retinal}. In turn, Jiang et al. in \cite{jiang2018retinal} proposed a supervised method to segment retinal blood vessel with the help of the fully convolutional network and transfer learning. In \cite{oliveira2018retinal}, Oliveria et al. proposed an approach that combined stationary wavelet transform with a multiscale fully connected network (FCN) to segment the different structures of blood vessels in retinal images. To improve the segmentation performance, Soomro et al. \cite{soomro2017boosting} proposed a method using deep conventional neural networks along with hysteresis threshold method for precise detection of the vessels. This method was able to detect smaller vessel structures compared to the previous state-of-the-art methods. 

In this paper, we propose a generative adversarial network (GAN) combined with spatial pyramid pooling and factorization networks to segment the blood vessels from the fundus images. GAN comprises of two successive networks: generator and discriminator. The generator network learns to map the input fundus image to the segmented image. This network includes three modules: spatial pyramid pooling to extract features on different scales, kernel factorization to reduce computational complexities and squeeze excitation block to enhance features representation in spatial and channel dimensions. In turn, the discriminator network of our adversarial framework is a CNN based classifier with an additional squeeze excitation block, where squeeze enhances the feature representation of the final convolution layers of the discriminator. Thus, our contributions are:
\begin{itemize}
	\item Pre-processing --- A pre-processing stage is proposed that converts a given colored eye fundus image to gray scale and enhances its sharpness to increase the segmentation possibility of thin vessels.
    \item Adversarial learning for segmentation --- An adversarial framework is proposed for retinal blood vessel segmentation. The generator network is formulated by using spatial pyramid pooling, kernel factorization, and squeeze excitation block for generating fine detailed vessel segmentation results. In turn, the discriminator network is modified by adding a squeeze excitation block in order to improve its classification accuracy.
    \item Post-processing--- A post-processing stage is proposed that removes the small pixelated binary noise using morphological operations. (for adversarial framework diagram, see suppl. A.1)
    \item Robustness --- The proposed blood vessel segmentation method is tested on two eye fundus image datasets, DRIVE and STARE,  to show its robustness. It yields comparable segmentation results to the state-of-the-art.
\end{itemize}

\vspace{-2mm}
\section{Proposed Methodology}
\begin{figure}[!t]
\centering
\includegraphics[scale=.45, width=\linewidth]{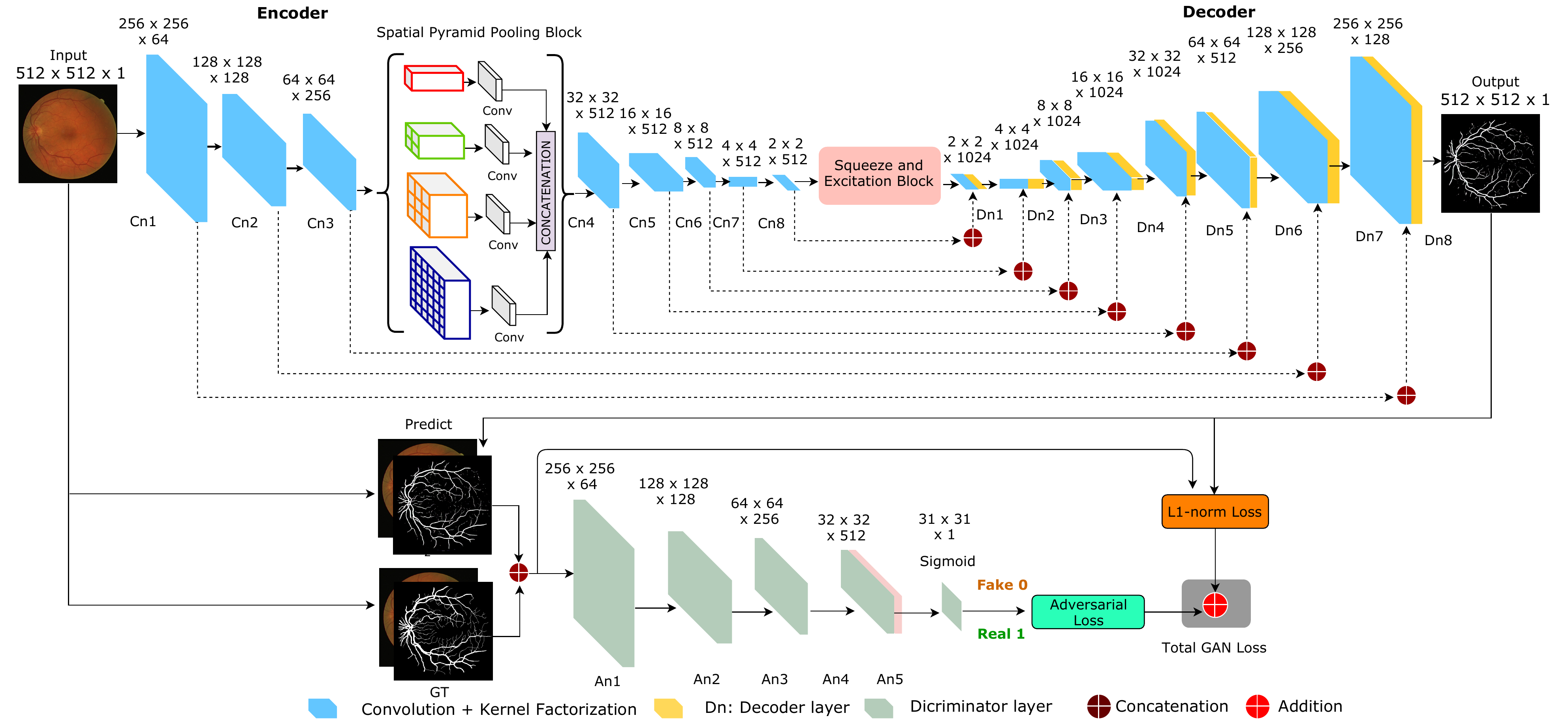}
\caption{The architecture of the proposed model. }
\label{fig1:architecture}
\end{figure}

\vspace{-3mm}
In this work, we propose an adversarial learning based retinal blood vessels segmentation method, which includes generator and discriminator networks. The generator network comprises an encoder and a decoder layers.  As shown in Figure \ref{fig1:architecture}, the encoder and decoder include 8 sequential layers ($C_n$ refers to an encoder layer and $D_n$ refers to a decoder layer). We  divide the proposed methodology into the following main concepts: generator network,  discriminator network,  a spatial pyramid pooling (SPP) \cite{he2015spatial}, a squeeze and  excitation block \cite{hu2018squeeze}, and residual 1D kernel factorization \cite {romera2018erfnet}.

\noindent\textbf{Generator}:
We have added a residual 1D kernel factorization block to each encoder and decoder layer. To gain more spatial information, an SPP block used in between the encoder layer $C_{n3}$ and $C_{n4}$. We have used four pool scales: $1$, $2$, $3$ and $6$. This pool scale helps to extract features in different scales that can learn large and small vessels from an input image and after concatenation of all these valuable features passed to the next encoder layer. The SPP layer works on each feature map independently. Additionally, at the end of the last encoding layer, we have used a squeeze and excitation block to adaptively recalibrates channel-wise feature responses by explicitly modeling interdependencies between channels. Besides, we added Leaky-ReLU  with a slope of 0.2 as an activation function after the first 7 layers of the encoder and ReLU after $C_{n8}$. To avoid the over-fitting problem batch normalization has been used after $C_{n1}$. We have been sequentially applied a kernel size of $4\times4$  with stride $2$ and padding $1$ at all convolutional layer of the encoder.\\
The decoder has a series of deconvolutional layers with kernel and stride size of $4\times4$  with stride $2$ respectively. To avoid  overfitting, we used batch normalization and dropout (rate = 0.5) in $D_{n1}$, $D_{n2}$ and $D_{n3}$. We added the ReLU activation function after each layer of the decoder. We have added a skip connection in between encoder and decoder layer where each convolutional layer has been concatenated to its corresponding deconvolutional layer. In  $D_{n8}$, the non-linear activation function \textit{tanh} has been used to outline the segmented retinal blood vessels lesion into a binary mask.\\
To optimize the proposed model, we have used a two different loss function into a generator network: binary cross entropy (BCE) and L1-norm Loss. The loss of the generator network can be defined as follows:

\begin{equation}
\ell_{Gen}(G, D) =  \mathbb{E}_{x,y,z}(-\log(D(x, G(x,z))))+ \lambda \mathbb{E}_{x,y,z}(\ell_{L1}(y, G(x,z)))
\end{equation}

Consider $x$ is a given retinal fundus image with blood vessels, $y$ is the ground truth mask, $G(x, z)$ and $D(x, G(x,z))$ are the outputs of the generator and the discriminator, respectively.

\noindent\textbf{Discriminator}: It comprises five convolutional layers with stride = 2 except $A_{n4}$ and $A_{n5}$ along with kernel size = 4. The input of the discriminator is the concatenation of the retinal fundus image and a binary mask marking the blood vessels region, where the vessels can either be the ground truth or the one predicted by the generator network. In $A_{n4}$, added a squeeze and excitation block to provide to automatically re-balance the impact of its output high level features. The output of discriminator network is followed by a sigmoid activation function to classify the input retinal fundus image mask as real or fake. The loss function of the discriminator $D$ can be explained as follows:
\begin{equation}
\label{Ldis}
\ell_{Dis}(G, D) = \mathbb{E}_{x,y,z}(-\log(D(x, y))) + \mathbb{E}_{x,y,z}(-\log(1-D(x, G(x,z))))
\end{equation}
The optimizer will fit $D$ to maximize the loss values for ground truth masks (by minimizing $-\log(D(x, y))$) and minimize the loss values for generated vessels (by minimizing $-\log(1-D(x, G(x, z))$). These terms calculate the BCE loss using both masks, assuming that the expected class for ground truth and generated masks is $1$ and $0$, respectively. The $G$ and $D$ are optimized concurrently,  one step for both networks at each iteration, where $G$ tries to generate a accurate vessels segmentation and $D$ learns how to discriminate fake masks from real ones.\\
\noindent\textbf{Kernel Factorized Block}: To reduce the computation complexity, suppress noise and increase the efficiency of the retinal blood vessels model residual 1-D kernel factorization has been used. This 1-D kernel factorized techniques helps to best support their learning performance and ability during training the proposed model. This block has used a kernel $1 \times 3$ and $3 \times 1$ with the dropout rate 0.3. There are many dilation rates has been applied in each convolutional and deconvolutional layer. The $C_{n1}$ and $C_{n2}$ used dilation rate 2, $C_{n3}$ and $C_{n4}$ used 4, $C_{n5}$ and $C_{n6}$ used 8, last two encoding layers $C_{n7}$ and $C_{n8}$ have employed a 16 dilation rate and similarly to the decoder. \\
Let $\textbf{W}\in\mathbb{R}^{C\times d^h\times d^v\times F}$ represent the weights of a classical 2D convolutional layer, where $C$ and $F$ refer to the number of input and output planes (feature map), respectively. Similarly, let $b\in\mathbb{R}^{F}$ be the vector representing the bias term for each filter,  $d^h\times d^v$ the kernel size of each feature map (typically $d^h \equiv d^v\equiv d$), and $\textbf{f}^\textbf{i} \in\mathbb{R}^{d^{h} \times d^{v}}$ the $i^{th}$ kernel in the layer. It is possible to rewrite $\textbf{f}^\textbf{i}$ by relaxing the rank-1 constraint as a linear combination of 1D filters as $\textbf{f}^\textbf{i}=\sum_{k=1}^{K} \sigma_{k}^{i} \Bar{v}_{k}^{i}\big( \Bar{h}_{k}^{i}\big)^T$, 
where $\Bar{v}_{k}^{i}$ and $\big(\Bar{h}_{k}^{i}\big)^T$ are vectors of length $d$, $\sigma_{k}^{i}$ is a scalar weight and $K$ is a rank of $\textbf{f}^\textbf{i}$. Thus, the $i^{th}$ output of the decomposed layer, $a_{i}^{1}$ can be written as a function of its input $a_{*}^{0}$ as
\begin {equation}
a_{i}^{1} = \varphi\bigg( b_{i}^{h} + \sum_{l=1}^{L} \Bar{h}_{il}^{T} * \bigg[\varphi\bigg(b_{l}^{v} + \sum_{c=1}^{C} \Bar{v}_{lc} * a_{c}^{0} \bigg)\bigg]\bigg)
\end{equation}

\noindent where $\varphi(.)$ represent the non-linearity of the 1D decomposed filters, which can be implemented with ReLU.



\section{Experimental Results}

\subsection{Experimental setup}
\noindent
\textbf{Dataset} In this paper, we have tested and compared our method with the state-of-the-art methods using two public eye blood vessel challenge datasets DRIVE \cite{niemeijer2004comparative} and STARE\cite{hoover1998locating}. The DRIVE dataset contains 20 training and 20 test images, in turn, STARE consists of only 20 images; therefore, we have manually splitted it into 10 training and 10 test images.

\noindent\textbf{Pre-processing}
We first converted RGB color space image to gray scale. After that contrast is enhanced by applying contrast limited adaptive histogram equalization (CLAHE) followed by the edge enhancement using edge sharpening. If $N$ and $M$ represent length and width of the given RGB image then the tiles for CLAHE are set to $\frac{N}{50} \times \frac{M}{50}$ with histogram bins $n=512$, whereas the radius and strength of the sharpening filter are $r=2.5$ and $s=3$, respectively. The difference of Gaussian is then applied to extract the edge enhanced gray scale image, where the $\sigma=10$ is the standard deviation of the Gaussian kernel.
We have augmented the data by using rotation, flipping, translation and gamma improvement to make sure there is no over and under fitting during the network training. After applying the data augmentation, we increase the number of images from 20 to 400 for DRIVE dataset and 10 to 200 for STARE dataset.
\\ \noindent \textbf{Model optimization}
The hyperparameters of the model were empirically tuned. We have experimented different optimizers, such as SGD, AdaGrad, Adadelta, RMSProp, and Adam with different learning rates. We obtained the best results with Adam optimizer ($\beta_1$= 0.5, $\beta_2$= 0.999  and learning rate =0.0002 with batch size 2). The L1-norm loss weighting factor $\lambda$  was set to 100. Both generator and discriminator are trained from scratch for 100 epochs.
\\ \noindent \textbf{Post-processing}
We first applied an average filter with a kernel size of $3\times3$ to regularize the segmentation result from the generator network. Then we connect three blocks of edge sharpening and image regularization block, where the radius and strength of the sharpening filter are $r=1$ and $s=1.15$. In turn, a circle type regularization kernel is used with a filter strength of $s=0.8$. Finally, a morphological operation called area open with a structural element of size $s_e=10$ is used to remove the small speckles of noise (small binary objects).
\vspace{-5mm}
\begin{table}[!h]
	\centering
	\caption{Quantitative comparison of proposed method results with the state-of-the-art methods using both DRIVE and STARE datasets}
	\label{table1:quantitative_comparison}
	\scalebox{0.8}{
		\begin{tabular}{|c|c|c|c|c|c|c|c|c|c|c|}
			\hline
			\multirow{2}{*}{\textcolor{blue}{\textbf{Methods}}} & \multicolumn{5}{c|}{\textcolor{blue}{\textbf{DRIVE}}}                   & \multicolumn{5}{c|}{\textcolor{blue}{\textbf{STARE}}}                   \\ \cline{2-11} 
			& AUC    & F1-score & Sen    & Spe    & Acc    & AUC    & F1-score & Sen    & Spe    & Acc    \\ \hline
			FCN \cite{oliveira2018retinal}            &  0.9748      & -          & 0.6706        & \textbf{0.9916}       & 0.9505       & 0.9846       &  -        & 0.8453       & 0.9726        & 0.9597       \\ \hline
			UNet \cite{ronneberger2015u}                   &  0.8622      & 0.7125         & 0.7003      & 0.9850        & 0.9404        & 0.9377       & 0.7006         & 0.7087       & 0.9783        & 0.9277       \\ \hline
			cGAN \cite{isola2017image}                    & 0.9314       & 0.7707         & 0.7258       & 0.9853       & 0.9623       & 0.9561       & 0.7452         & 0.7448       & 0.9810       & 0.9624       \\ \hline
			Hu et al. \cite{hu2018retinal}         & 0.9759 & --       & 0.7772 & 0.9793 & 0.9533 & 0.9751 & --       & 0.7543 & 0.9814 & 0.9632 \\ \hline
			Jiang et al. \cite{jiang2018retinal}      & 0.9810 & --       & 0.7540 & 0.9825 & 0.9624 & \textbf{0.9900} & --       & \textbf{0.8352} & \textbf{0.9846} & 0.9734 \\ \hline
			Soomro et al. \cite{soomro2017boosting}     & 0.8310 & --       & 0.7460 & 0.9170 & 0.9460 & 0.8350 & --       & 0.7480 & 0.9220 & 0.9480 \\ \hline
			\textcolor{red}{Proposed}                 & \textcolor{red}{\textbf{0.9890}}       & \textcolor{red}{\textbf{0.8003}}   & \textcolor{red}{\textbf{0.7851}}        & \textcolor{red}{0.9834}       & \textcolor{red}{\textbf{0.9659}}       &  \textcolor{red}{0.9860}      & \textcolor{red}{\textbf{0.7710}}         & \textcolor{red}{0.7634}        & \textcolor{red}{0.9830}       &   \textcolor{red}{\textbf{0.9812}}     \\ \hline
		\end{tabular}}
	\end{table}
	\begin{figure}[!t]
		\centering
		\includegraphics[width=0.8\textwidth, height=0.18\textheight]{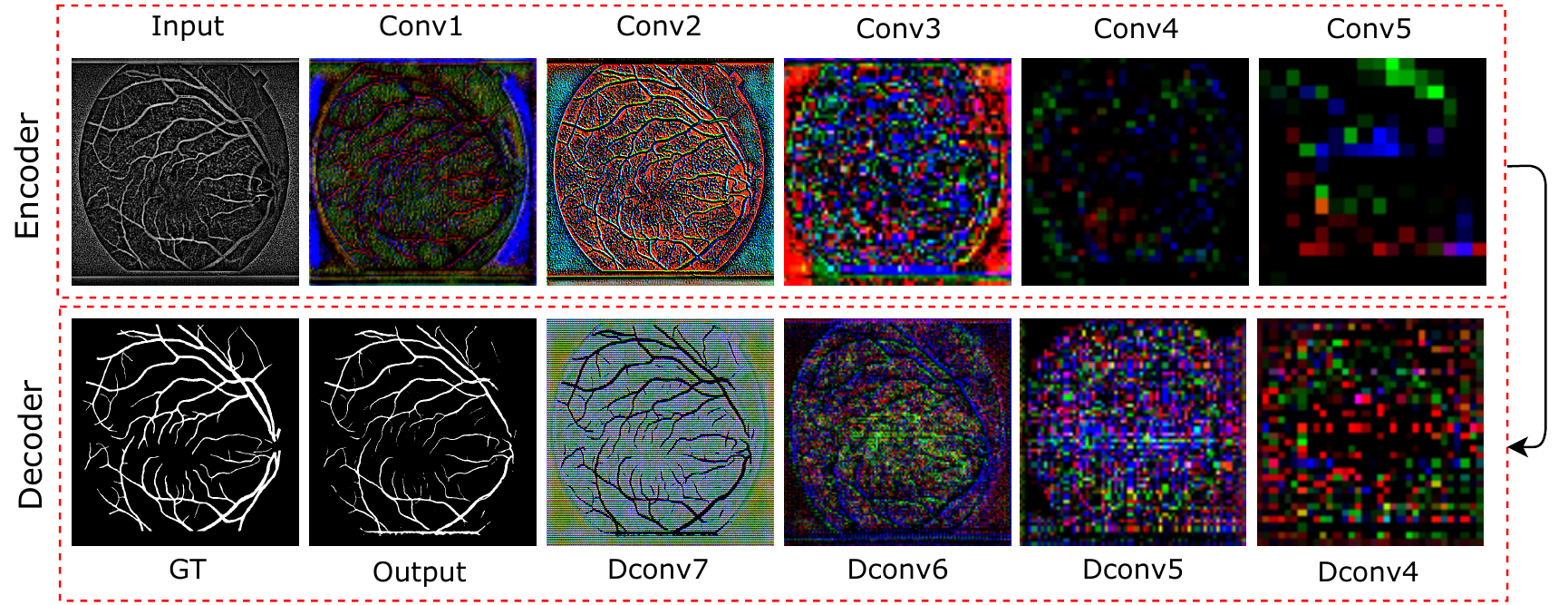}
		\caption{Visualizing encoder and decoder layers. }
		\label{fig2:visualization}
		\vspace{-2mm}
	\end{figure}
	
	\subsection{Results and Discussion}
	In this section, the results are computed using the proposed method and compared with the state-of-the-art eye blood vessel segmentation methods both quantitatively and qualitatively. Table~\ref{table1:quantitative_comparison} shows a quantitative comparison between the proposed method and the state-of-the art segmentation methods: FCN\cite{oliveira2018retinal}, UNet\cite{ronneberger2015u}, cGAN\cite{isola2017image}, Hu et al. \cite{hu2018retinal}, Jiang et al. \cite{jiang2018retinal}  and Soomro et al. \cite{soomro2017boosting} using both DRIVE and STARE dataset for retinal blood vessel segmentation. For the comparison, different similarity metrics are used such as area under the curve (AUC), F1-score, sensitivity, specificity and accuracy. It shows that for the DRIVE dataset, the proposed method yields the best values for all the metrics except specificity. It yields AUC, F1-score, specificity and accuracy of 0.9890, 0.8003, 0.7851 and 0.9659, respectively. Where the AUC is approx $1\%$ more than of the second best Jiang et al. method, F1-score $3\%$ more than of the second best cGAN, sensitivity approx $1\%$ more than of the second best Hu et al. method and the accuracy is only $0.3\%$ more than of the second best Jiang et al. method. In turn, the proposed method yields a specificity of 0.9834, which is approx. $1\%$ less than of the FCN.
	
	\indent For the STARE dataset, the proposed method yields the best values for F1-score and accuracy metrics. It yields the best F1-score of 0.7710, which is $2.58\%$ higher than the second best cGAN. Moreover, it yields the best accuracy of 0.9812, which is approx $1\%$ more than the second best Jiang et al. method. On the other hand, the proposed method yields AUC, sensitivity and specificity of 0.9860, 0.7634 and 0.9830, which are $0.4\%$, $7.18\%$ and $0.16\%$ less than the second best Jiang et al. method.
	
	\indent Figure~\ref{fig2:visualization} shows a layer by layer visualization of our model, which shows how results are being generated by the generator network. The top row (from left to right) shows how pre-processed image is passed from the first to fifth layer of the encoder layers and at each layer after being processed by different filters. The bottom row (from right to left) shows how the output of the encoder network is being decoded from the fourth to last layer of decoder network. Note that we have not added last three encoding and first three decoding layers in our visualization as these layers do not contain any detailed structures.  
	
	\begin{figure}[!t]
		\centering
		\includegraphics[width=1\textwidth, height=0.22\textheight]{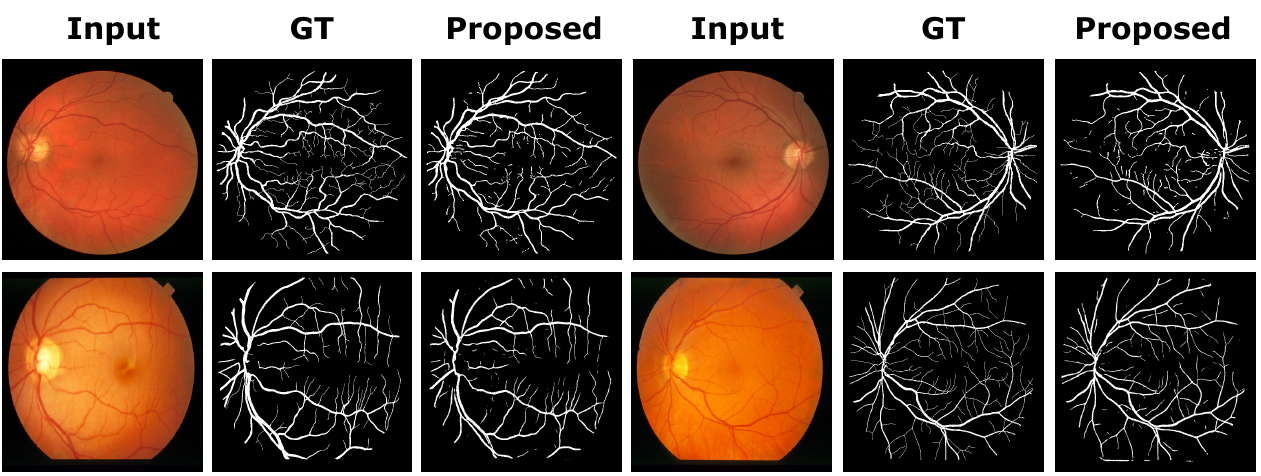}
		\caption{The segmentation output of the proposed model. }
		\label{fig3:segmentation}
	\end{figure}
	
	\begin{figure}[!t]
		\centering
		\includegraphics[width=1\textwidth, height=0.22\textheight]{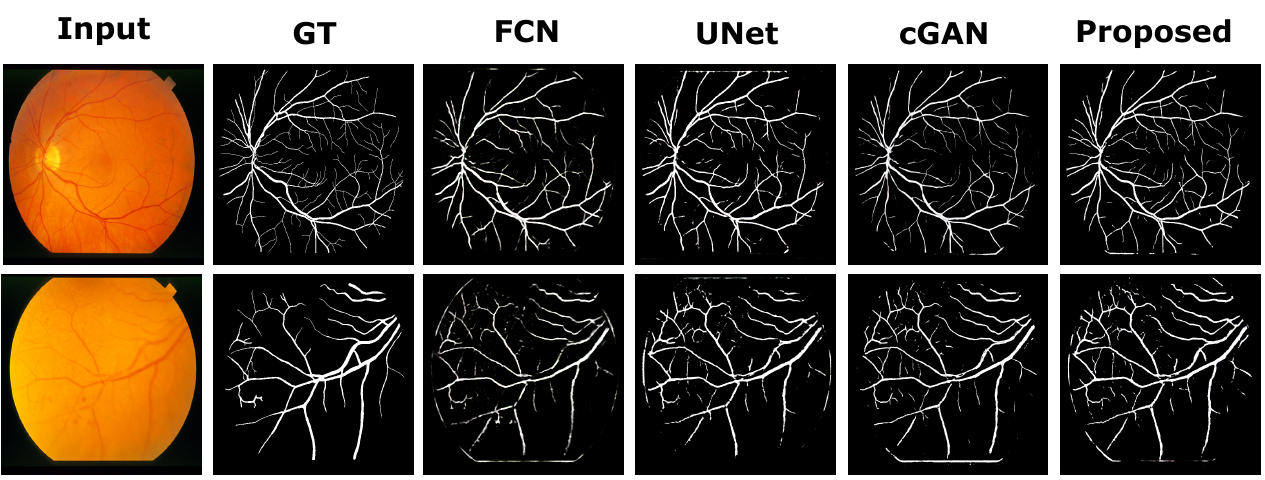}
		\caption{Comparing the proposed model with the state-of-the-art methods. }
		\label{fig4:qualitative_comparison}
	\end{figure}
	Figure~\ref{fig3:segmentation} shows a retinal vessel segmentation results using the proposed method. It shows that the proposed method is able to properly segment thicker vessels, which similar as its respective ground truth. However, in some cases there are few discontinuities observed in the segmented thin vessels that is the limitations of the proposed method. Figure~\ref{fig4:qualitative_comparison} shows a qualitative comparison between the proposed, FCN \cite{oliveira2018retinal}, UNet \cite{ronneberger2015u} and cGAN methods\cite{isola2017image}. The visual comparison shows that the proposed method yields the best segmentation results than the compared methods. It is able to segment fine details and the thin vessels, whereas the other methods failed to do so. As discussed earlier, the proposed model also observes some discontinuities in the segmented thin vessel; however, it is better than the compared methods which even miss those vessels. Note that all experiments are performed in the same work environment using same datasets (for more results and experiments, see suppl. A.2).
	\vspace{-2mm}
	\section{Conclusions}
	\vspace{-2mm}
	In this paper, an efficient retinal blood vessel segmentation method was proposed a generative adversarial network. The proposed generative adversarial network consists of generator and discriminator networks. The generator network is formulated by using three modules: spatial pyramid pooling, kernel factorization, and squeeze excitation block for generating exceptionally detailed vessel segmentation results. In turn, the discriminator includes a squeeze excitation block to enhance the feature representation. The proposed model outperformed the state-of-the-art in terms of accuracy, F1-score and AUC. However, it still is unable to segment a few thin vessels properly. In future work, we will improve our model to target thin vessels by introducing a deep learning model based on an edge enhancement mechanism.
	\vspace{-3mm}
	\bibliographystyle{splncs04}
	\bibliography{biblography} 

\begin{thebibliography}{10}
\providecommand{\url}[1]{\texttt{#1}}
\providecommand{\urlprefix}{URL }
\providecommand{\doi}[1]{https://doi.org/#1}

\bibitem{he2015spatial}
He, K., Zhang, X., Ren, S., Sun, J.: Spatial pyramid pooling in deep
  convolutional networks for visual recognition. IEEE Trans Pattern Anal Mach
  Intell  \textbf{37}(9),  1904--1916 (2015)

\bibitem{hoover1998locating}
Hoover, A., Kouznetsova, V., Goldbaum, M.: Locating blood vessels in retinal
  images by piece-wise threshold probing of a matched filter response. In:
  Proc. of the AMIA Symposium. p.~931. American Medical Informatics Association
  (1998)

\bibitem{hu2018squeeze}
Hu, J., Shen, L., Sun, G.: Squeeze-and-excitation networks. In: Proc. of the
  IEEE conference on computer vision and pattern recognition. pp. 7132--7141
  (2018)

\bibitem{hu2018retinal}
Hu, K., Zhang, Z., Niu, X., Zhang, C., Xiao, F., Gao, X.: Retinal vessel
  segmentation of color fundus images using multiscale convolutional neural
  network with an improved cross-entropy loss function. Neurocomputing
  \textbf{309},  179--191 (2018)

\bibitem{isola2017image}
Isola, P., Zhu, J.Y., Zhou, T., Efros, A.A.: Image-to-image translation with
  conditional adversarial networks. In: Proc. of the IEEE conference on
  computer vision and pattern recognition. pp. 1125--1134 (2017)

\bibitem{jiang2018retinal}
Jiang, Z., Zhang, H., Wang, Y., Ko, S.B.: Retinal blood vessel segmentation
  using fully convolutional network with transfer learning. Comput Med Imaging
  Graph  \textbf{68},  1--15 (2018)

\bibitem{macgillivray2014retinal}
MacGillivray, T., Trucco, E., Cameron, J., Dhillon, B., Houston, J., Van~Beek,
  E.: Retinal imaging as a source of biomarkers for diagnosis, characterization
  and prognosis of chronic illness or long-term conditions. The British journal
  of radiology  \textbf{87}(1040),  20130832 (2014)

\bibitem{niemeijer2004comparative}
Niemeijer, M., Staal, J., van Ginneken, B., Loog, M., Abramoff, M.D.:
  Comparative study of retinal vessel segmentation methods on a new publicly
  available database. In: Medical imaging 2004: image processing. vol.~5370,
  pp. 648--657. International Society for Optics and Photonics (2004)

\bibitem{oliveira2018retinal}
Oliveira, A., Pereira, S., Silva, C.A.: Retinal vessel segmentation based on
  fully convolutional neural networks. Expert Syst Appl  \textbf{112},
  229--242 (2018)

\bibitem{romera2018erfnet}
Romera, E., Alvarez, J.M., Bergasa, L.M., Arroyo, R.: Erfnet: Efficient
  residual factorized convnet for real-time semantic segmentation. IEEE T
  INTELL TRANSP  \textbf{19}(1),  263--272 (2018)

\bibitem{ronneberger2015u}
Ronneberger, O., Fischer, P., Brox, T.: U-net: Convolutional networks for
  biomedical image segmentation. In: International Conference on Medical image
  computing and computer-assisted intervention. pp. 234--241. Springer (2015)

\bibitem{soomro2017boosting}
Soomro, T.A., Afifi, A.J., Gao, J., Hellwich, O., Khan, M.A., Paul, M., Zheng,
  L.: Boosting sensitivity of a retinal vessel segmentation algorithm with
  convolutional neural network. In: 2017 International Conference on Digital
  Image Computing: Techniques and Applications (DICTA). pp.~1--8. IEEE (2017)

\bibitem{vostatek2017performance}
Vostatek, P., Claridge, E., Uusitalo, H., Hauta-Kasari, M., F{\"a}lt, P.,
  Lensu, L.: Performance comparison of publicly available retinal blood vessel
  segmentation methods. Comput Med Imaging Graph  \textbf{55},  2--12 (2017)

\end{thebibliography}

\end{document}